\documentclass[aps,prl,groupedaddress,amsfonts,twoside,twocolumn,amssymb,final]{revtex4}

\usepackage{amsmath}
\usepackage{amsfonts}
\usepackage{amssymb}
\usepackage{graphicx}

\def\idty{\openone}

\def\abs#1{\vert#1\vert}
\def\norm#1{\Vert#1\Vert}
\def\ketbra#1#2{\vert#1\rangle\langle#2\vert}
\def\braket#1#2{\langle#1\vert#2\rangle}
\def\ket#1{\vert#1\rangle}
\def\bra#1{\langle#1\vert}
\def\tr{{\rm tr}}

\def\HH{{\cal H}}\def\B{{\cal B}}
\let\BB\B

\def\HS{{\cal L}^2} 
\def\Jam#1{\widetilde{#1}} 
\def\Braket#1#2{\langle\langle#1\vert#2\rangle\rangle} 

\def\Kett{\bigr\rangle\mkern-5mu\big\rangle}
\def\Braa{\big\langle\mkern-5mu\bigl\langle}

\def\Braket#1#2{\Braa#1\big\vert#2\Kett}
\def\Ketbra#1#2{\bigr\vert#1\Kett\Braa#2\bigl\vert}

\DeclareMathOperator{\id}{id}

\begin{document}
\title{Iterative Optimization of Quantum Error Correcting Codes}
\author{M. Reimpell, R.~F. Werner }
\affiliation{Institut f\"{u}r Mathematische Physik, TU-Braunschweig, Mendelssohnstra{\ss}e 3,
  D-38106 Braunschweig, Germany}

\begin{abstract}
We introduce a convergent iterative algorithm for finding the
optimal coding and decoding operations for an arbitrary noisy
quantum channel. This algorithm does not require any error
syndrome to be corrected completely, and hence also finds codes
outside the usual Knill-Laflamme definition of error correcting
codes. The iteration is shown to improve the figure of merit
``channel fidelity'' in every step.
\end{abstract}

\maketitle

\section{Introduction}
From the beginning of the development of quantum information theory, it
was recognized that without suitable error correcting procedures the
fantastic promises of this new discipline, such as the exponential
speedup in Shor's algorithm, or the possibility of long range quantum
communication and secure key exchange would never be realizable.
Therefore, the development of the first error correcting codes \cite{lit:shor,
lit:steane}
and the subsequent more systematic theory by Knill and Laflamme
\cite{lit:knillLaflamme} were crucial achievements. It became clear that
although naive
classical ideas, like redundant transmission and majority rule decisions
on the outputs, are ruled out by the no-cloning theorem, techniques from
classical coding theory (e.g., additive codes) could be used to construct
good quantum codes as well. The quantum codes constructed in this way
share with their classical counterparts the combinatorial/algebraic
flavor. They are designed to correct a certain finite dimensional
subspace of errors, such as errors occurring on only a small
number of the parallel channels employed. If the space of corrected
errors is suitably chosen, such codes can also be used to correct generic
small errors, i.e., one can show that any channel close to the ideal
channel can be corrected with small overhead \cite{lit:barnumKnillNielsen}.

\begin{figure}
  \begin{center}
    \includegraphics*[width=6.2cm]{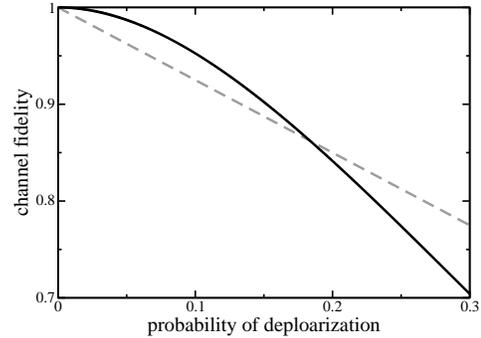}
    \caption{\label{fig:fbcPart}Fidelity of the five bit code applied to
    $5$-fold tensor product of the depolarizing qubit channel (solid line)
    compared to the fidelity of the depolarizing channel (dashed line).}
  \end{center}
\end{figure}

However, for errors of fixed finite size it is not at all clear
that the special form of Knill-Laflamme codes allows the most
efficient error correction. Alternative codes might not correct
any error completely, but in exchange might improve correction of
the errors ignored by the Knill-Laflamme codes, resulting in an
improved overall performance. Consider, for example, the famous
five bit code~\cite{lit:laflammeMiquelPazZurek,
lit:bennettDiVincenzoSmolinWootters}, applied to the five-fold
tensor product of a depolarizing qubit channel with a certain
depolarization probability $p$. Figure~\ref{fig:fbcPart} shows the
fidelity achieved by this code as a function of $p$, together with
the same parameter without any correction. For $p> 1-\sqrt{2/3}
\approx 18 \%$ the performance of the five bit code is actually
worse than doing no correction operation at all. It seems
implausible that the best code should jump from using all qubits
to using only one at the crossover point, which suggests looking
for better codes in that area.

In this Letter we develop a method which allows us to search
numerically for an optimal code adapted to arbitrary noise. Thus,
for any given noisy channel $T$ (not necessarily a product
$T=S^{\otimes n}$ of channels operating independently on $n$
smaller systems), we look for an {\it encoding channel} $E$ and a
{\it decoding channel} $D$, with suitable domain and range, such
that $DTE$ comes as close as possible to the ideal channel on a
fixed $d$-level system. In contrast to Knill-Laflamme theory we
make no assumptions on the coding and decoding channels $E$ and
$D$. The basis of the method is an iteration by which either $E$
or $D$ is changed, such that fidelity is improved in each step.
The results are related to Knill-Laflamme theory as follows:

\begin{enumerate}
 \item Surprisingly, the codes in Fig.\ref{fig:fbcPart} turn out to
be optimal already: up to the critical value of the depolarization
probability the five-bit code is optimal, and beyond that the best
way of using up to five bit encodings is to do nothing. However,
this has little bearing on general channels, since the
depolarizing channels are highly symmetric.
 \item The encoding operation comes out to be an isometry even on
random channels. This is a basic feature of Knill-Laflamme theory.
 \item Sometimes the Knill-Laflamme theory applies, but not, as is
usually done, to the correction of localized errors.
Instead, certain non-localized errors are corrected. Such
instances are also known in the decoherence-free subspace approach
\cite{lit:Zanardi}, and are reliably found by our method, since we do not use a
tensor product decomposition of $T$ in the first place.
\item Sometimes Knill-Laflamme theory fails entirely in the sense
that no error at all is corrected completely: although $E,D$
optimally correct a channel $T$, there might be no channel $T'$
such that $DT'E$ is a multiple of the identity.
\end{enumerate}

\section{Overview of the method}
Quantum information theory describes computation in terms of
preparation, processing and measurements. We consider only finite
dimensional quantum systems, i.e., systems whose observable
algebra is of the form $\BB(\HH)$, the linear operators on a
finite dimensional Hilbert space $\HH$. The quantum states, which
physically describe the preparation process are given by density
operators $\rho$ in $\BB(\HH)$. Measurements are given by
selfadjoint operators on $\HH$, or, more generally by positive
operator-valued measures.  Processing operations, e.g. the free
time evolution, a computation or a noisy transmission, are
described by channels. These can either be considered as a
modification of all subsequent measurements (Heisenberg picture),
or as a modification of the preparation (Schr\"odinger picture).
In this article we choose the latter option, i.e., channels are
mathematically given by completely positive trace preserving
operators $S:\BB(\HH_1)\to\BB(\HH_2)$, where $\HH_1$ is the
Hilbert space of the input systems, and $\HH_2$ describes the
output systems, and $S(\rho)$ is the state obtained by sending the
input state $\rho$ through the channel. The encoding and decoding
operations of an error correction scheme are also channels in this
sense, with appropriate choices of input and output Hilbert
spaces. Every channel $S$ has a Kraus representation
$S(\rho)=\sum_is_i\rho s_i^*$, with $s_i\mathpunct:\HH_1\to\HH_2$,
and $\sum_is_i^*s_i=\idty$. When $\HH_1=\HH_2$, the 'noisiness' of
$S$ is, loosely speaking, its distance from the ideal channel.
There are many different ways of expressing this quantitatively.
In this Letter we use a special case of Schumacher's Entanglement
Fidelity~\cite{lit:schumacher}, the {\it channel fidelity}. It is
defined as
\begin{equation} \label{eqn:channelFidelity}
\begin{split}
F_C(S) & = \Big\langle\Omega\Big\vert \bigl(\id \otimes S\bigr)(\ketbra
              {\Omega}{\Omega}) \Big\vert\Omega\Bigr\rangle \\
       & = (\dim\HH_1)^{-2}\sum_i\abs{\tr(s_i)}^2\;,
\end{split}
\end{equation}%
where $\ket \Omega = (\dim \HH_1)^{(-1/2)}\sum_k \ket {kk}$ is the
standard maximally entangled unit vector in $\HH_1 \otimes \HH_1$ 
and $\id$ is the identity channel on $\BB(\HH_1)$.
This quantity is $1$ if and only if the channel is ideal, and is
directly related to the mean fidelity for pure input states
\cite{lit:horodeckiHorodeckiHorodecki}.

The problem of finding an optimal code for a given channel
$T\mathpunct:\BB(\HH_1)\to\BB(\HH_2)$ is now the construction of
an encoding channel $E\mathpunct:\BB(\HH_0)\to\BB(\HH_1)$ and a
decoding channel $D\mathpunct:\BB(\HH_2)\to\BB(\HH_0)$ such that
$F_C(DTE)$ becomes maximal. This is always a fairly high
dimensional search problem.  For example, if $\HH_0$ is a single
qubit, and $T$ is the five-fold tensor power of a given noisy
channel ($\dim\HH_1=32$), i.e., the case considered in
figure~\ref{fig:fbcPart}, the description of $D$ and $E$ together
requires some $7000$ parameters. General purpose optimization
routines will usually choke on this, and there only is a chance if
special properties of this variational problem can be brought to
bear.

What we use in the present Letter is that the functionals $E\mapsto
F_C(DTE)$ and $D\mapsto F_C(DTE)$ are both linear, and take
positive values on completely positive operators. The iteration
procedure described in the next section finds a maximum of any
functional with these properties. The overall maximization then
proceeds see-saw fashion, by applying the iteration first with a
fixed random $E$, optimizing the fidelity over $D$, then fixing
$D$ and optimizing $E$, and so on. Since every step of the
iteration is proved to increase fidelity and each stable fixed
point of the single iteration is a global maximum, this procedure
is guaranteed to find at least a locally optimal pair of encoder
and decoder.

\section{The basic iteration}
The iteration we consider is a close relative of the {\it power
method} for finding the eigenvector for the largest eigenvalue of
a positive semi-definite matrix $A$. This method starts with a
random unit vector $\phi_0$, and each step consists of applying
$A$, and normalizing, i.e., $\phi_{n+1}=A\phi_n/\norm{A\phi_n}$.
It is easy to see that the convergence of this algorithm is
exponential with a rate determined by the gap to the next largest
eigenvalue. Moreover, the inequality
\begin{equation}\label{ineqA}
  \frac{\braket{\phi}{A^3\phi}}{\braket{\phi}{A^2\phi}}
    \geq \frac{\braket{\phi}{A\phi}}{\braket{\phi}{\phi}}   \;,
\end{equation}
which is valid for any positive semi-definite Hilbert space operator $A$,
shows that convergence is {\it monotone}, in the sense that
$\braket{\phi_n}{A\phi_n}$ is a non-decreasing sequence.

Suppose now that we want to find a channel $S$ maximizing a linear
objective functional $f$, which is defined on arbitrary operators
$S:\BB(\HH_1)\to\BB(\HH_2)$, and positive on all completely
positive maps. Note that $f(S)$ is bilinear in the Kraus operators
$s_i$ of $S$. So in a sense we will presently make precise,
the objective functional $f$ is analogous to the matrix element
$\braket{\phi_S}{\widetilde F\phi_S}$ of a positive operator
associated with $f$, where the vector $\phi_S$ corresponds to the
set of Kraus operators $s_i$. In our iteration we apply the
operator $\widetilde F$ to each Kraus operator and get a modified
completely positive map. This will not be a channel, because it is
not trace preserving. Hence we have to include a normalization
step. Since the normalization of a completely positive map is
given by an operator (not a scalar) one cannot simply ``divide by
the normalization''. We show, however, how to do the normalization
in such a way that the desirable features of the power method do
carry over.

Let us now make these ideas precise. By $\HS(\HH_1,\HH_2)$ we
denote the space of Hilbert Schmidt operators from $\HH_1$ to
$\HH_2$ with scalar product $\Braket xy=\tr(x^*y)$. Then if
$\ket\mu$, $\mu=1,\ldots,\dim\HH_1$ denotes the vectors of a basis
of $\HH_1$, we associate with any map
$S\mathpunct:\B(\HH_1)\to\B(\HH_2)$ an operator
 $\Jam S\in \B\bigl(\HS(\HH_1,\HH_2)\bigr)$
by
 \begin{equation}\label{Jam1}
  \Jam S(x)=\sum_{\mu\nu}S(\ketbra \mu\nu)\;x\ketbra \nu\mu  \;,
\end{equation}
In fact, this is just a reshuffling of matrix elements since $\bra{a}
S(\ketbra\mu\nu) \ket{b} = \bra{a} \Jam{S}(\ketbra{b}{\nu})\ket{\mu}$.
The key feature of the correspondence $S\leftrightarrow\Jam S$, also
known as the Jamiolkowski duality,  is that $S$ is completely positive if
and only if $\Jam S$ is a positive semi-definite operator on the Hilbert
space $\HS(\HH_1,\HH_2)$. Indeed, the Kraus decomposition of $S$
translates directly into
\begin{equation}\label{jams-kraus}
  \Jam S= \sum_i \Ketbra{s_i}{s_i}\;,
\end{equation}
 and the operators with such a representation are precisely the positive
semi-definite operators on $\HS(\HH_1,\HH_2)$.

The objective functional $f(S)$ can now be written in terms of $\Jam S$,
and thus becomes a positive linear functional on the positive operators
on $\HS(\HH_1,\HH_2)$. But such functionals are themselves given by
positive operators: there is a positive semi-definite $\Jam F$ such that
\begin{equation}\label{eqn:dualObjective}
  f(S) = \tr(\Jam F \Jam S)
  =\sum_i \Braket{s_i}{\Jam F s_i}\;,
\end{equation}
where at the second equality we have inserted Eq.~(\ref{jams-kraus}).

In each iteration we define a new completely positive
map $S'$ by applying $\Jam F$ to each $s_i$, i.e., $s_i'=\Jam F(s_i)$, or
\begin{equation}\label{jamsprime}
  \Jam S'=\sum_i\Jam F\Ketbra{s_i}{s_i}\Jam F
         =\Jam F\Jam S\Jam F\;.
\end{equation}
Clearly, $S'$ is usually not trace preserving. Instead, we have
$\tr(S'(\rho))=\tr(M\rho)$, where
\begin{equation}\label{mprime}
  M=\sum_i(s_i')^*s_i'\;.
\end{equation}
In order to normalize the channel we therefore multiply each $s_i'$ with
the suitable power of $M$: if $M$ is non-singular, we set
$t_i=s_i'M^{-1/2}$, so $\sum_it_i^*t_i=\idty$. These will be the Kraus
operators of the next iterate $S_{+}$, i.e., the overall iteration step
is
\begin{equation}\label{tt}
  S\mapsto S_{+}, \quad S_{+}(\rho)=S'(M^{-1/2}\rho M^{-1/2})\;,
\end{equation}
with $S'$ determined by Eq.~(\ref{jamsprime}). In the applications below
$M$ is always invertible. But when $M$ is singular, we can still take
$M^{-1/2}$ as the pseudo-inverse, and the channel $S_{+}$ becomes
normalized to a projection, i.e., it is trace preserving only for input
density matrices on the support subspace of $M$ and annihilates density
operators supported on the complement.

The properties of this iteration resemble those of the power
method (which is, in fact, the special case $\dim\HH_2=1$). Most
importantly, one gets an improvement of the objective functional
in every step: $f(S_{+})\geq f(S)$. The proof is based on
inequality (\ref{ineqA}), for an operator $A$ depending on the
normalization correction $M$, which hence changes in every step.
As for the power method, there may be non-maximal fixed points of
the iteration, corresponding to non-maximal eigenvalues of $A$.
However, they are all unstable: a small random perturbation of
such a fixed point is sufficient to get the iteration going again,
finding strictly higher $f(S)$. Therefore, testing the stability
of any fixed point found is included into the general algorithm.

We have proved that this ``stabilized'' iteration does converge to
the global maximum, provided that the initial number of Kraus
operators is sufficiently large to allow representation of
arbitrary channels for the given dimensions. In that case
convexity guarantees that there are no sub-optimal local maxima,
and the linear stability analysis of the iteration shows that all
stable points are indeed local maxima. Note that our iteration
without stabilization never increases the number of Kraus
operators, so we can also find local maxima with such a
constraint, e.g., the constraint that encoding uses only one
isometry.

\section{Application to Quantum Error Correction}

As mentioned above, we will optimize the overall fidelity of the
corrected channel $F_C(DTE)$ by alternately fixing the decoding
$D$ and optimizing the encoder $E$ by the iteration method, and
fixing $E$ and optimizing $D$. Since both kinds of steps increase
fidelity, this procedure converges to an optimum. All results
reported below were computed by starting from various random
initial confi\-gurations. The iteration was stopped when the
gain of fidelity was below some threshold.%

\subsection{Depolarizing Channel}
The procedure is applied to the depolarizing qubit channel with parameter
$p$, i.e.,
\begin{equation}\label{depolChannel}
   T_p(\rho) =  p\;\tr (\rho) \frac12\idty + (1- p)\rho.
\end{equation}%
For $0 \leq p \leq 1$ this channel totally depolarizes the input
system with probability $p$ and leaves the input system untouched
with probability $(1-p)$. The importance of this channel lies in
its role as the worst case (the most mixed channel), whenever only
a lower bound on the fidelity of a channel is known. The
correction scheme for the depolarizing channel will then correct
all such channels, to at least the same fidelity, even if further
details are unknown. However, due to its high symmetry this
channel is rather special (see Subsect.~C below).

\begin{center}
\begin{figure}
    \includegraphics*[width=6.2cm]{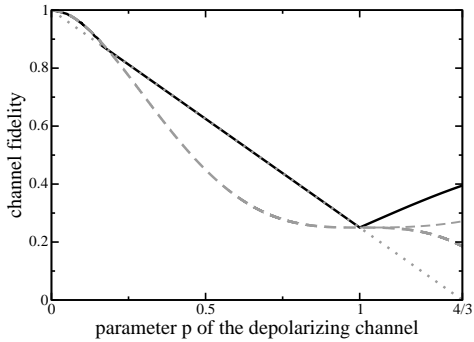}
    \caption{\label{fig:depolarization}
Comparison of the channel fidelity of no error correction (dotted
line), five bit code (dashed lines) and the iteration (solid line)
applied to the $5$-fold tensor product of depolarizing channel
with parameter $p$. For $p>1$ also the fidelity for five-bit
encoder combined with optimized decoder is shown.}
\end{figure}%
\end{center}

We will look at the fivefold tensor power of the depolarizing
channel, since for fewer copies of the channel the optimal
correction strategy turns out to do no correction at all, i. e.,
to copy the input to one of the output qubits and discard the
rest. For five bits we have the standard five-bit stabilizer
code~\cite{lit:laflammeMiquelPazZurek, lit:bennettDiVincenzoSmolinWootters},
which we denote by $(E_5,D_5)$. Its performance, given by the polynomial%
\begin{equation}\label{5biterr}
F_C(D_{5}T_p^{\otimes 5}E_{5}) = 1 -\frac{45}{8}p^2
+\frac{75}{8}p^3 -\frac{45}{8}p^4 +\frac{9}{8}p^5 \;,
\end{equation}%
is shown in Figure~\ref{fig:depolarization}. Surprisingly,
the optimal codes determined by our method fall exactly on the
known lines: the five-bit code up to the cross-over point $p =
1-\sqrt{2/3} \approx0.18$, and doing nothing for the range up to
$p=1$.  This is very surprising in view of the fact that the five
bit code is not at all designed to give good results for large
errors, but only to eliminate the linear term in~\eqref{5biterr}.

\subsection{New Codes near the Universal Not}
Note that values $p>1$ in equation~\eqref{depolChannel} are also
admissible, as it defines a completely positive map for all $0\leq
p\leq4/3$. For $p > 1$ the channel correspond to a mixing of the
totally depolarizing channel and the best possible approximation
to the ``universal not'' channel \cite{lit:unot}. In this range of
$p$ our method does lead to a new type of code. By this we mean
that in contrast to Knill-Laflamme theory {\it no error syndrome
is corrected}: there is no channel $T'$ such that $DT'E$ is a
multiple of the identity, whereas any corrected error syndrome in
the Knill-Laflamme theory would provide such $T'$. We establish
this result by our basic iteration, this time fixing $E,D$, and
considering $T'$ as the variable. On the other hand, by fixing
$E=E_5$ and $T$, one can also check that it is not sufficient to
just improve the decoder, and keep the five-bit-code encoding, as
suggested by the analogous classical case of three bit flip channels
with flip probability greater than $1/2$.

\subsection{Random Channels}
Moving away from the highly symmetric channels, we have considered
random channels generated either with independent uniformly distributed
entries, followed by normalization, or as convex combinations of such
channels with the identity. In either case one sees that, generically,
the optimized codes never correct a single syndrome. In contrast to the known
limitations of Knill-Laflamme codes, even for four encoding bits one often gets
an improvement of the fidelity. More precisely, the fidelity after coding tends
to increase (though often not by much) with every additional encoding qubit.
 
On the other hand, one feature of Knill-Laflamme theory is
typically shared by the optimized codes: the encoding $E$ is
isometric, i.e., it is given by a single Kraus operator. While it
is known that this choice is asymptotically optimal (it suffices
to get the same capacity as general encodings
\cite{lit:barnumKnillNielsen}), it is open whether it is also
optimal for every fixed noisy channel, as suggested by our random
search.

\section{Acknowledgement}
We thank Koenraad M. R. Audenaert and Markus Grassl for discussions. Funding
from Deutsche Forschungsgemeinschaft is gratefully acknowledged.

\end{document}